\documentclass[aip,jcp,amsmath,amssymb,reprint,floatfix]{revtex4-1}

\usepackage{graphicx}
\usepackage{dcolumn}
\usepackage{bm}

\usepackage[utf8]{inputenc}
\usepackage[T1]{fontenc}
\usepackage{mathptmx}
\usepackage{color}
\usepackage{hyperref}
\usepackage{tabularx}

\usepackage{pdfpages}
\usepackage{pgffor}

\newcommand{\ensl}{D\'epartement de Physique, \'Ecole Normale Sup\'erieure de Lyon,  46 All\'ee d'Italie, Lyon Cedex 07, France}
\newcommand{\ilm}{Univ Lyon, Univ Claude Bernard Lyon 1, CNRS, Institut Lumi\`ere Mati\`ere, F-69622, VILLEURBANNE, France}
\newcommand{\iuf}{Institut Universitaire de France (IUF)}

\makeatletter
\AtBeginDocument{\let\LS@rot\@undefined}
\makeatother

\begin{document}

\title{Molecular modeling of aqueous electrolytes at interfaces: effects of long-range dispersion forces and of ionic charge rescaling} 

\author{Guillaume Le Breton}
\affiliation{\ensl} 
\affiliation{\ilm}
\author{Laurent Joly}
\email{laurent.joly@univ-lyon1.fr}
\affiliation{\ilm}
\affiliation{\iuf}

\date{\today}

\begin{abstract}
Molecular dynamics simulations of aqueous electrolytes generally rely on empirical force fields, combining dispersion interactions -- described by a truncated Lennard-Jones (LJ) potential -- and electrostatic interactions -- described by a Coulomb potential computed with a long-range solver. Recently, force fields using rescaled ionic charges (electronic continuum correction, ECC), possibly complemented with rescaling of LJ parameters (electronic continuum correction rescaled, ECCR), have shown promising results in bulk, but their performance at interfaces has been less explored.  
Here we started by exploring the impact of the LJ potential truncation on the surface tension of a sodium chloride aqueous solution. We show a discrepancy between the numerical predictions for truncated LJ interactions with a large cutoff and for untruncated LJ interactions computed with a long-range solver, which can bias comparison of force field predictions with experiments. 
Using a long-range solver for LJ interactions, 
we then show that an ionic charge rescaling factor 
chosen to correct long-range electrostatic interactions in bulk also describes accurately image charge repulsion at the liquid-vapor interface, and that the rescaling of LJ parameters in ECCR models -- aimed at capturing local ion-ion and ion-water interactions in bulk -- also describes well the formation of an ionic double layer at the liquid-vapor interface.    
Overall, these results suggest that the molecular modeling of aqueous electrolytes at interfaces would benefit from using long-range solvers for dispersion forces, and from using ECCR models, where the charge rescaling factor should be chosen to correct long-range electrostatic interactions. 
\end{abstract}


\maketitle 

\section{Introduction}

Molecular dynamics (MD) is a very powerful tool to explore the structure and dynamics of aqueous electrolytes at the atomic scale. To simulate large systems over long times, empirical interaction potentials (force fields) are widely used. 
Liquid water is commonly described with 
rigid non-polarizable models \cite{Vega2011,Vega2015}, 
and some of them perform quite well. For instance, the SPC/E model \cite{spce} is rather good at reproducing the dielectric properties of water \cite{RamiReddy1989,Bonthuis2011,Schlaich2016}, and the TIP4P/2005 model \cite{tip4p_2005} reproduces accurately the structure and dynamics of water over a wide range of temperatures and pressures \cite{Vega2006,Pi2009,Rozmanov2012,Russo2014,Biddle2017,Guillaud2017,MonteroDeHijes2018}. 
However, non-polarizable models for ions in water are less successful at predicting the thermodynamics and dynamics of aqueous solutions \cite{Moucka2013,Moucka2013a,Nezbeda2016}. For instance, most non-polarizable models cannot reproduce even qualitatively the impact some salts have on water self-diffusion \cite{Kim2012,Ding2014}, while explicit inclusion of polarizability and/or charge transfer can improve the predictions \cite{Yao2014,Yao2015,Nguyen2018}.

In that context, new non-polarizable models have been developed based on a rescaling of the ionic charges \cite{Leontyev2003,Leontyev2009,Leontyev2011,Pegado2012,Kann2014,Kohagen2014,Kohagen2016,Benavides2017,Kroutil2017,Martinek2018,Bruce2018,Yue2019,Zeron2019,Jorge2019}, an approach often referred to as electronic continuum correction (ECC). Originally, the rescaling aims at implicitly describing electronic polarization, 
to improve the description of local ion-water and ion-ion interactions \cite{Leontyev2011}. But the rescaling can also compensate for the underestimated permittivity of water models  \cite{Kann2014}, and recover the correct long-range Coulombic interactions. 
Both motivations suggest different charge rescaling factors, i.e. $1/\sqrt{\varepsilon_\text{el}}$ (with $\varepsilon_\text{el}$ the electronic permittivity of the solvent) for the local argument \cite{Leontyev2011}, and $\sqrt{\epsilon_r/\epsilon_r^\text{exp}}$ (with $\epsilon_r$ the permittivity of the water model and $\epsilon_r^\text{exp}$ the experimental value) for the long-range argument \cite{Kann2014}. In practice, various rescaling factors have been chosen  \cite{Leontyev2003,Leontyev2009,Leontyev2011,Pegado2012,Kann2014,Kohagen2014,Kohagen2016,Benavides2017,Kroutil2017,Martinek2018,Bruce2018,Yue2019,Zeron2019,Jorge2019}, based on the expressions above or simply tuned to optimize the performance of the model.
Bare ECC, or ECC complemented with rescaling of the Lennard-Jones (LJ) parameters \cite{Kohagen2014,Kroutil2017} -- referred to as ECCR for 'electronic continuum correction rescaled', indeed provide improved predictions for the structure, dielectric permittivity and dynamics of bulk aqueous solutions~\cite{Kann2014,Renou2014a,Kohagen2014,Kohagen2016,Benavides2017,Bruce2018,Zeron2019,Yue2019,laage2019effect}.

However, ECC models have been less studied at interfaces~\cite{Vazdar2012,Neyt2014,Biriukov2018}. 
At the water-air interface, the ECC increases the surface affinity of ions and can create an ionic double layer \cite{Vazdar2012,Neyt2014}, in line with the predictions of polarizable force fields~\cite{vrbka2004propensity,DAuria2009,Neyt2013}. Yet the bare ECC overestimates the anionic surface affinity, an effect attributed to the abrupt change in the electronic part of the relative permittivity across the interface \cite{Vazdar2012}. Moreover, the experimentally observed linear increase of surface tension with respect to ionic concentration is not always  recovered \cite{Neyt2014}. 
ECC models can also be applied to liquid-solid interfaces, by rescaling the surface charges consistently with those of the electrolyte \cite{Biriukov2018}.

A crucial test of the force field performance at interfaces is to compare its prediction for the surface tension with experimental results~\cite{paul2003dynamics,dos2008consistency,alejandre2010surface,isele2012development,Netz2012,Vazdar2012,Neyt2014,Ghoufi2016,Benavides2017,obeidat2019new,ghoufi2019calculation}. 
With that regard, previous work has shown that the standard truncation of the Lennard-Jones (LJ) interaction potential can lead to large quantitative differences in the surface tension \cite{Sega2017}, or even to qualitatively different behaviors of liquids at interfaces \cite{Trokhymchuk1999,Valeriani2007,Caupin2008a,Alejandre2010,Evans2016,Fitzner2017}. Analytical tail corrections are commonly used for the surface tension \cite{Ghoufi2016}, but their implementation can be complex -- especially for electrolyte solutions, and there is no guarantee that the structure and dynamics of the interface are correctly predicted by truncated potentials. 
Alternatively, methods commonly used to compute untruncated Coulomb interactions by calculating the long-range part of the interaction in the Fourier space \cite{deserno1998mesh} can also be applied to LJ interactions \cite{isele2012development,isele2013reconsidering}. In particular, these approaches successfully predict liquid-vapor surface tension, without requiring a posteriori corrections \cite{isele2012development,isele2013reconsidering}. 

In that context, here we will investigate sequentially two important issues for the description  of aqueous electrolytes at the liquid-vapor interface, focusing on sodium chloride. First, we will explore the impact of LJ potential truncation on the liquid-vapor surface tension, and show the interest of using a long-range solver for LJ interactions. 
We will then use such a solver to explore the impact of charge rescaling, and identify the best choices to describe accurately the interfacial structure and surface tension of aqueous electrolytes.

\section{Systems and methods}

We will use the ECCR model by \citet{Benavides2017}, referred to as the Madrid model from the authors group's location.
This model is based on TIP4P/2005 water; the charge rescaling factor of $0.85$ is closer to the value suggested by the long-range argument ($0.86$) than by the local argument ($0.75$). 

\begin{figure}
\centering\includegraphics[width=0.8\linewidth]{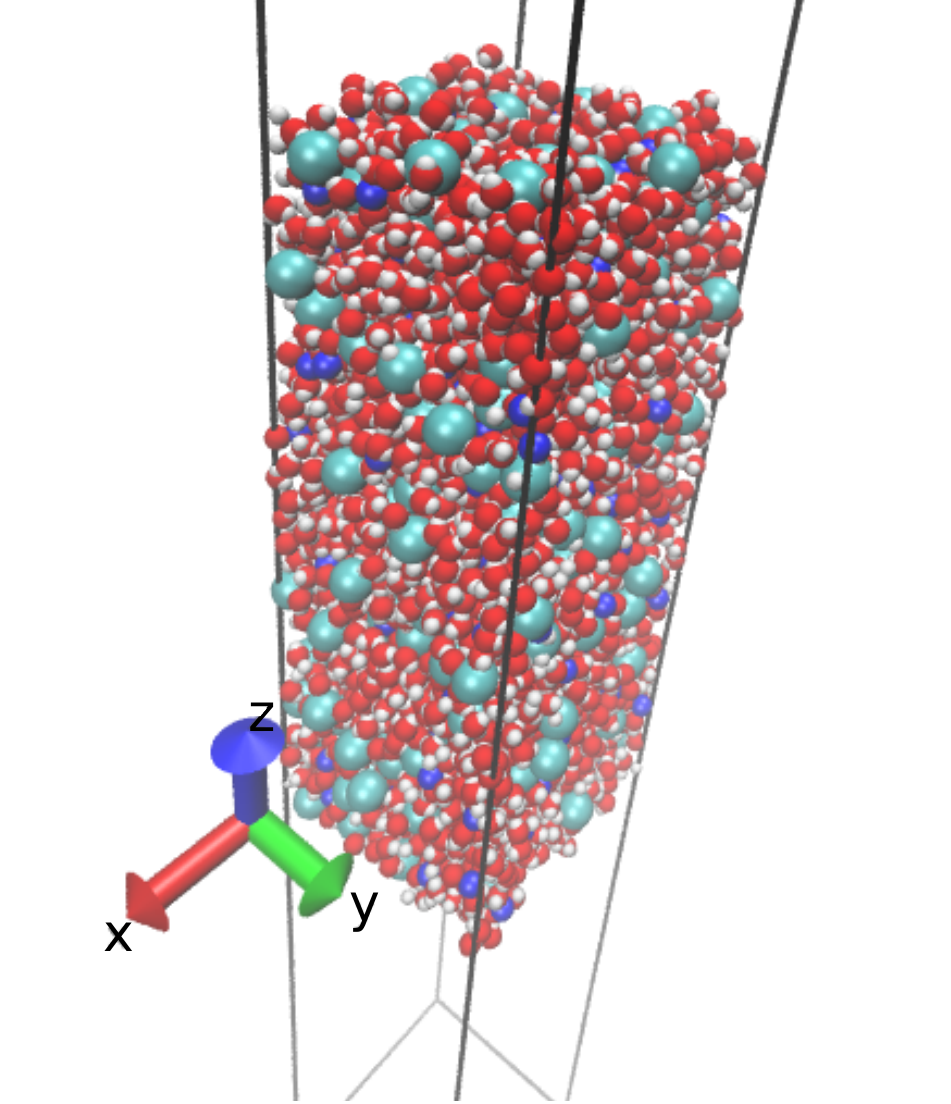}
\caption{\label{Fig: visualisation system} Snapshot of a typical system: 3500 water molecules, 128 Cl$^-$ and 128 Na$^+$ ions, corresponding to a 1.6\,mol/L system. The box size is $34 \times 34 \times 300$\,\AA{}$^3$, the system extension in the z-axis is around $90$\,\AA{}. Periodic boundary conditions are used in the 3 directions.}
\end{figure}

We simulated a liquid film (along the $x-y$ plane) illustrated in Fig.~\ref{Fig: visualisation system},
composed of 3500 water molecules. 
The initial systems were built by using MOLTEMPLATE \cite{MOLTEMPLATE}, and we used LAMMPS \cite{LAMMPS} to run the simulations.  
The tested NaCl concentrations were 0.1, 0.8, 1.6, 3.2 and 4.2\,mol/L. 
The total box size was $34.5 \times 34.5 \times 300$\,\AA{}$^3$, and the extension of the liquid phase in the $z$ direction was about $90$\,\AA{}. 
Periodic boundary conditions were used along the 3 directions. The vacuum gap in the $z$-direction was large enough to effectively remove interactions between the system and its periodic images in the $z$ direction. These simulation box values have been widely used in the literature and have been shown to be sufficient to prevent finite size effects \cite{Alejandre2010, Vega2007}. We also tested finite size effects, as detailed in the supplemental material (SM).  
We integrated the equations of motion using the velocity-Verlet algorithm, with a time step of 2\,fs.
Long-range Coulombic interactions were treated with the  particle-particle–particle-mesh (PPPM) method, a point-grid based Ewald method. 
Water molecules were held rigid using the SHAKE algorithm. 

The system was equilibrated during ca. 3\,ns, and the production run lasted for 100\,ns. 
We calculated the surface tension $\gamma$ from the difference of normal and tangential pressure, as detailed in the SM:  
\begin{equation}
\gamma = \frac{L_z}{2} \left[ p_z - \frac{1}{2}(p_x + p_y) \right] , 
\end{equation} 
where $p_i$ is the average pressure along direction $i = x,y, z$, and $L_z$ is the total box size along the $z$ direction normal to the two interfaces. 
Experimentally, the surface tension increases linearly with respect to the ionic bulk concentration; the surface tension gain between 0.1\,mol/L and 4.2\,mol/L is around 7\,mN/m for NaCl at room temperature \cite{washburn1928international,allen2009shedding}.

\section{Effect of long-range dispersion interactions}

As discussed in the introduction, the standard procedure of truncating LJ interactions at a distance of ca. 1\,nm, and possibly applying analytical tail corrections, has been challenged recently for heterogeneous systems \cite{Sega2017,Trokhymchuk1999,Valeriani2007,Alejandre2010,Fitzner2017}. 
Therefore, we have tested here the impact of the truncation procedure. 
First, we have computed the surface tension of pure SPC/E and TIP4P/2005 water using various cutoffs (note that we used a simple cutoff scheme without any smoothing) and an Ewald based method -- PPPM, presented in  Refs.~\citenum{isele2012development,isele2013reconsidering} -- to treat the LJ interaction.
As detailed in the SM, for truncated LJ interactions, the surface tension seems to converge at high cutoff value. For pure water, the interfacial density profile is well approximated by an hyperbolic tangent shape, for which analytical tail corrections can be derived \cite{Chapela1977,Blokhuis1995}, providing satisfying results since the corrected surface tension reaches a plateau. For both SPC/E and TIP4P/2005 water, using the PPPM method fixes the LJ cutoff dependence. Moreover, for the pure water system, results obtained using PPPM and using cutoffs with tail corrections matched quantitatively. 
Still, the surface tensions obtained with the PPPM method are ca. 2\,mN/m larger than the ones obtained using the largest cutoff (17\,\AA{}) without tail correction. This highlights the fact that the long range part of the LJ potential -- naturally taken into account through the PPPM method -- has a significant impact and even a large cutoff misses some relevant interaction for this heterogeneous system. 


\begin{figure}
\centering
\includegraphics[width=0.99\linewidth]{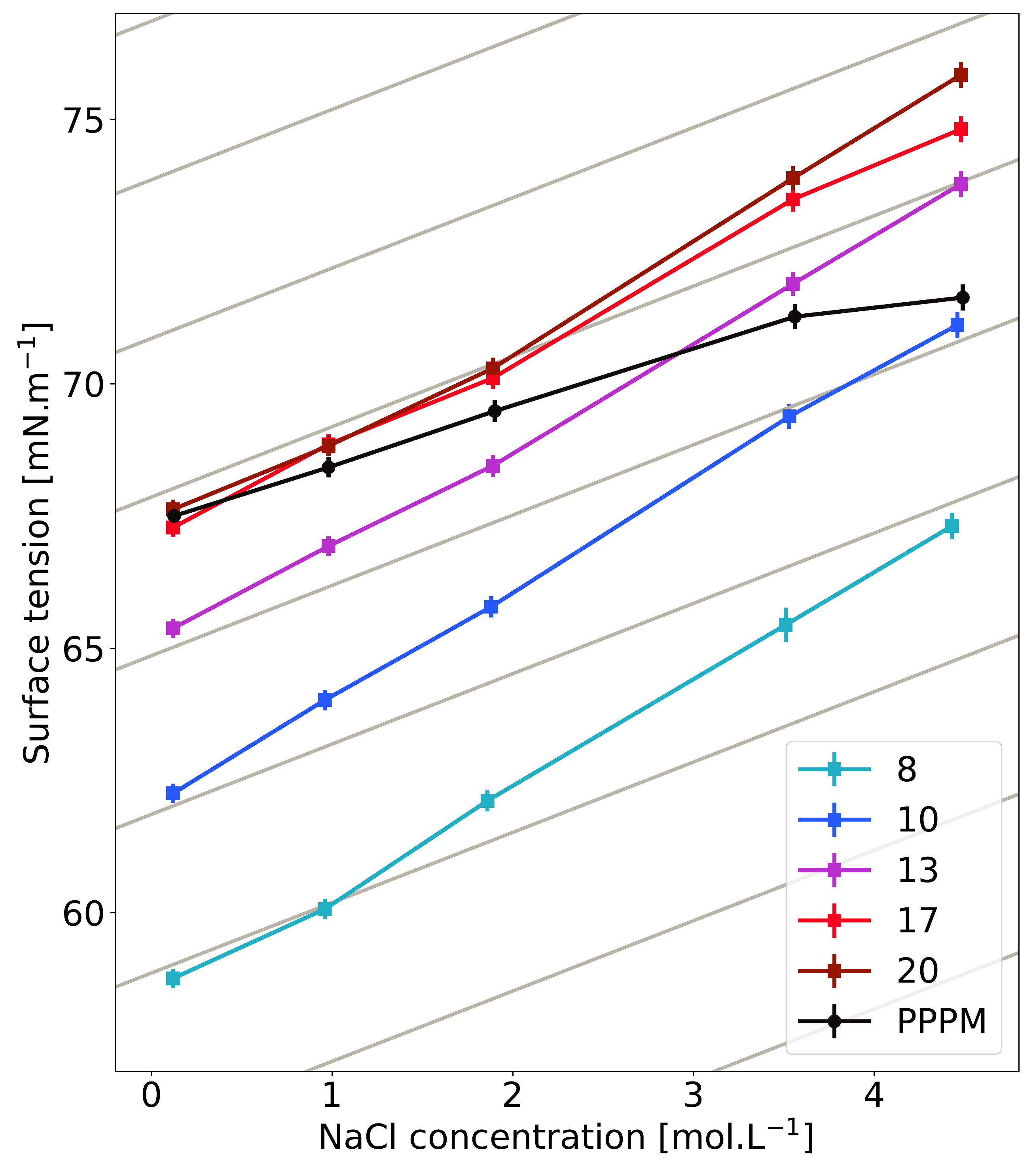}
\caption{\label{Fig: figure 2} Surface tension of the Madrid model of aqueous NaCl as a function of NaCl concentration, using truncated LJ interactions with different cutoffs (squares), or untruncated LJ interactions with the PPPM method (black circles). Since no analytical correction is used, the surface tension for pure water increases with increasing cutoff. The error bars correspond to a 95\,\% confidence level -- see the SM for more details. For comparison, the tilted gray lines array indicates the experimental gain \cite{washburn1928international}. 
}
\end{figure}

We then tested the effect of the long-range part of dispersion interactions in the presence of salt, by comparing the results obtained with the cutoff and the PPPM methods, using the Madrid model of NaCl in water. 
Note that for an aqueous electrolyte solution, no simple tail correction can be written due to the complex ion distribution at the interface, so that here we only considered the raw simulation results.
At low salt concentration, the surface tension obtained with the cutoff method is smaller than the one using the PPPM method, see Fig.~\ref{Fig: figure 2}. This is consistent with the results obtained for pure water, see Fig.~1 of the SM. 
As shown by the force field developers \cite{Benavides2017}, the surface tension increases smoothly with the salt concentration, in contrast with previous results obtained with another ECC model \cite{Neyt2014}.  
Surprisingly, when the salt concentration increases, the increase in surface tension is higher with the cutoff method than with the PPPM method, see Fig.~\ref{Fig: figure 2}. 
Importantly, this effect could bias comparisons of the surface tension dependency on salt concentration with experiments, which are commonly used as a test of the quality of aqueous electrolyte force fields. Here for instance, cutoff simulations predict higher increases of the surface tension with respect to experiments while PPPM predicts lower one, so that the model could be validated or not depending on how long-range LJ interactions are treated. 
The PPPM method has the advantage to be consistent for any system and does not require any post-processing treatment. Therefore, this approach will be used in the following.

\section{Effect of charge rescaling}

\begin{figure*}
\centering
\includegraphics[width=0.9\linewidth]{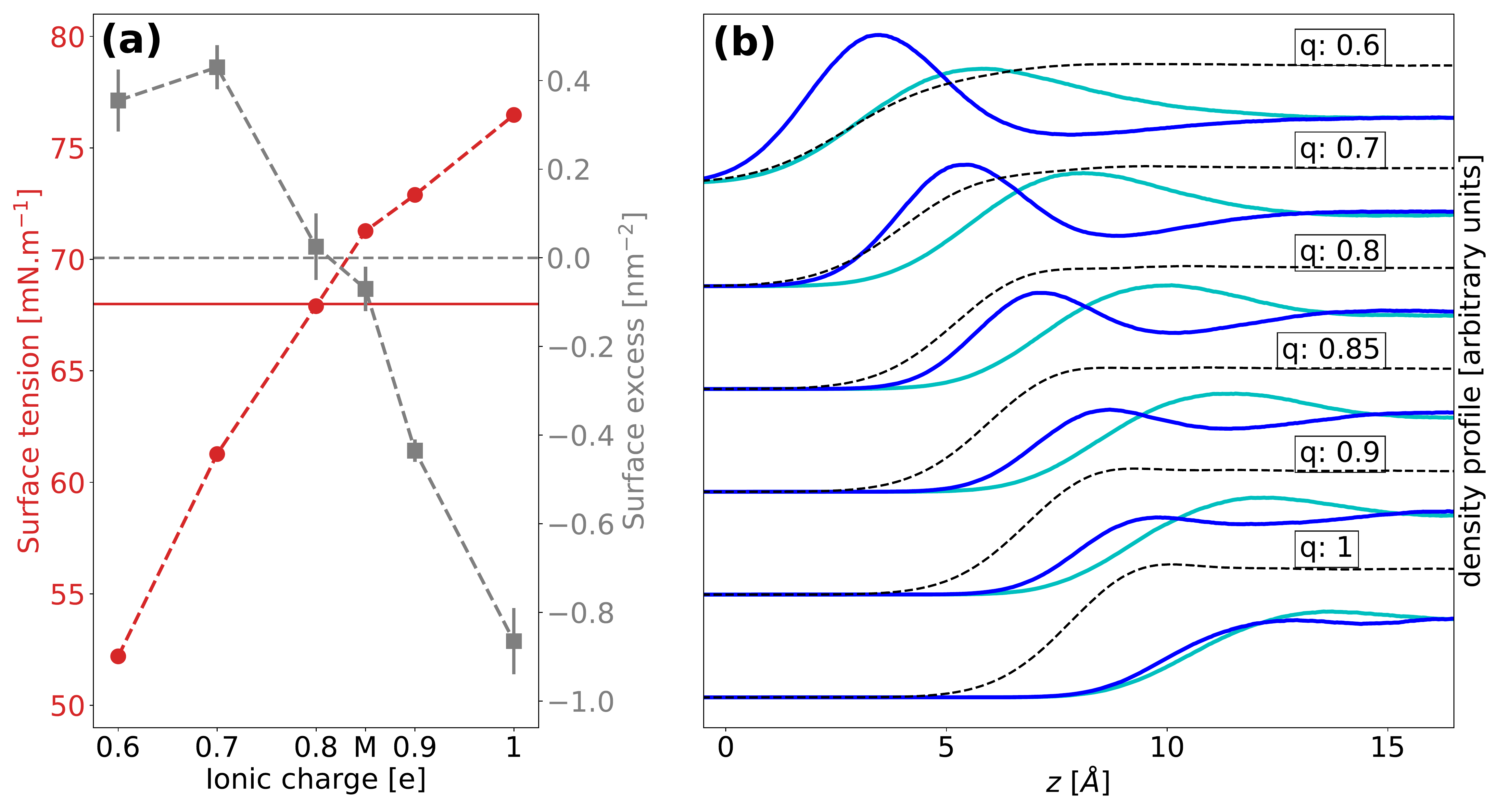}
\caption{\label{Fig: Charge effect Surface tension} Effect of ionic charge rescaling on the surface tension, surface excess, and density profiles of a 3.2\,mol/L NaCl solution with Madrid's VdW parameters and TIP4P/2005 water model. \textbf{(a)} Left axis and red circles: surface tension versus ionic charge; the horizontal red full line represents the computed surface tension of pure water, 68\,mN/m; according to the experimental surface tension gain as a function of the concentration, the expected value for the surface tension at 3.2\,mol/L is 72-73\,mN/m; note that the error bars are within the symbols. Right axis, gray squares: surface excess (SE) versus ionic charge; the gray dashed line is the zero SE value; experiments indicate a negative SE -- since the surface tension gain is positive with respect to the ionic concentration, see Eq.~\eqref{eq:gamma};  the `M' on the abscissa stands for the orinal Madrid parameters: ionic charge of $\pm$~0.85. \textbf{(b)} Density profiles of water (black dashed lines), Cl$^{-}$ (dark blue lines), and Na$^+$ (light cyan lines), for different ionic charges. 
Water density profiles have been normalized to appear on the same scale as the ionic density profiles. 
}
\end{figure*}

A good starting point to explore the effect of charge rescaling at an interface is to fix the LJ parameters (here we are using those of the Madrid model), 
and modify only the charges at a given concentration, 3.2\,mol/L. 
The resulting surface tensions, surface excess (SE) and density profiles are shown in Fig.~\ref{Fig: Charge effect Surface tension}, which highlights the dramatic impact of ionic charge. 
For a $\pm \ e$ charge, Cl$^{-}$ and Na$^{+}$ are identically depleted from the interface and fully solvated, 
see bottom part of Figure~\ref{Fig: Charge effect Surface tension}b. 
These observations are consistent with other non-polarisable MD simulations using fully charged ions \cite{Bhatt2004,huang2008aqueous}.
When the charge is decreased, 
the ions increasingly adsorb at the interface, with adsorption peaks growing and moving toward the surface. The ions also organize in a 'double layer', where the Cl$^{-}$ adsorption peak is closer to the surface than the Na$^{+}$ peak. 
This ionic double layer is an expected feature of the NaCl water-air system, which is retrieved using QMMM simulations \cite{Jungwirth2001} or polarizable classical MD \cite{vrbka2004propensity,DAuria2009,Neyt2013}.

In order to quantify this structural behavior, we have computed the surface excess (SE), denoted $\Gamma_s$, for the different charge rescaling (see the SM for more details), which is plotted in Fig.~\ref{Fig: Charge effect Surface tension}a. 
The SE is related to the evolution of the surface tension $\gamma$ with respect to the solute activity $a_s$ through Gibbs’ thermodynamic theory of interfaces:   
\begin{equation}\label{eq:gamma}
\Gamma_s = -\frac{1}{R T} \left( \frac{\partial \gamma}{\partial \ln a_s} \right)_T , 
\end{equation}
with $R$ the gas constant and $T$ the temperature. 
For NaCl, $\gamma$ increases with the salt concentration, so that $\Gamma_s$ is negative, with an experimental value on the order of $-0.5$\,nm$^{-2}$, see Ref.~\citenum{DAuria2009}. 

Classical polarizable models \cite{DAuria2009,Vazdar2012,ishiyama2007molecular} provide correct SE values, while standard non-polarizable ones predict too negative values \cite{DAuria2009}. In Ref.~\citenum{Vazdar2012}, an ECC approach with a  $\pm \ 0.75e$ ionic charge (chosen based on short range arguments), predicts a positive SE of 0.06\,nm$^{-2}$ for a concentration of ca. 0.8\,mol/L, and thus a negative surface tension gain with respect to ionic concentration. 
Our results are fully consistent with the previous ones: for a ionic charge of 0.6-0.8$\,e$, the SE is positive and the surface tension gain is negative, while for a $\pm e$ charge the SE is very  negative and leads to the highest surface tension gain. 

The original Madrid force field, with a ionic charge of $\pm 0.85 \,e$, 
predicts values very close to the experimental ones. 
Of course one can attribute this success to the additional work done for the VdW parametrization since the Madrid model belongs to the ECCR class. As pointed out in Ref.~\citenum{Biriukov2018}, rescaling the VdW parameters helps obtaining better results also at interfaces. But, for these kind of systems, we would like to argue that the rescaling factor should be chosen based on the long-range argument (i.e., correcting long-range Coulomb interactions), as is the case for the Madrid model, and not on the short-range one. 
To that aim, we will use a simple mean field model inspired by Ref.~\citenum{huang2008aqueous}. 
Ionic density profiles $\rho_{\pm} (z)$ at the liquid-vapor interface follow a Boltzmann distribution: 
$\rho_{\pm} (z) = \rho_0 \exp \left\{ -\beta U_{\pm}(z) \right\}$, where $\rho_0$ is the bulk ionic density, $\beta = 1/(k_\text{B} T)$, and $U_{\pm}(z)$ the potential felt by the ion. 
This potential can be decomposed as follows: 
$U_{\pm}(z) = \pm eV(z) + U_{\pm}^{\text{solvation}}(z) + U_{\pm}^{\text{image}}(z)$, 
where $V(z)$ is the electrostatic potential, $U_{\pm}^{\text{solvation}}(z)$ represents the interaction with the solvent, and $U_{\pm}^{\text{image}}(z)$ is an image charge potential acting on ions near the dielectric interface located at $z=0$, as described by Onsager-Samaras theory \cite{onsager1934surface}: 
\begin{equation}\label{eq:os}
U_{\pm}^{\text{image}}(z) = \left( \frac{\epsilon_r-1}{\epsilon_r+1} \right) \frac{q^2 \exp \left[ -2  z/\lambda_D \right]}{16 \pi \epsilon_0 \epsilon_r z} , 
\end{equation}
with $\epsilon_0$ the vacuum permittivity, $\epsilon_r$ the solvant relative permittivity, $q$ the ionic charge, and $\lambda_D = \sqrt{\epsilon_0 \epsilon_r k_B T/(2q^2 \rho_0)}$ the Debye length. Here it is important to note that the solvant relative permittivity involved is the bulk value far from the interface. This potential identically pushes both cations and anions inside water. 
From Eq.~\eqref{eq:os}, it is clear that choosing the rescaled ionic charge to correct for the permittivity of the water model in bulk, i.e., $q = e \sqrt{\epsilon_r/\epsilon_r^\text{exp}}$, will also adequately correct the image potential. Indeed, because water (and water models) have a very large $\epsilon_r$, the error induced by the water model in the prefactor $(\epsilon_r-1)/(\epsilon_r+1)$ -- not corrected by the rescaling procedure -- is minimal. For instance, for TIP4P/2005 water at room temperature, $\epsilon_r \approx 60$ while $\epsilon_r^\text{exp} \approx 80$, corresponding to an error in the prefactor of around 1\,\% only. 
This explains why a rescaling of $\pm \ 0.85 \, e$ for the TIP4P/2005 water model provides a structure -- and in particular a SE -- close to the expected one at the interface, while this rescaling has been originally designed for bulk systems. 
In contrast, when no rescaling is appplied, the image potential is too strong, which lead to very negative SE, and with a rescaling of $\pm \ 0.75 \, e$, the image potential is too weak, letting the ions go too far toward the vapor side -- leading to too positive SE. 

Beyond the image potential, the detailed ion distribution will be controlled by the other terms in the potential felt by the ions, $\pm eV(z) + U_{\pm}^{\text{solvation}}(z)$. With that regard, our simulations show that using an ECCR approach, as done for the Madrid FF, captures correctly the distribution predicted by polarized FFs -- and in particular the formation of a double layer. 
To understand this result, one should note that in practice, charges and LJ interaction parameters in ECCR models are tuned empirically to accurately describe the local environment of ions, i.e. first neighbor ion-ion and ion-water interactions. While the parametrization is performed in bulk, one can expect that the first neighbor interactions should also be fairly described at interfaces, even though the ion hydration shells are different.

\section{Conclusions}

We have shown that similar surface tensions are found using the cutoff and the PPPM method for pure water and dilute salts -- reaching quantitative agreement when adding tail corrections. In contrast, at high salt concentration, where no simple tail correction can be written due to the complex ion distribution at the interface, the cutoff method leads to a larger surface tension increase than the one obtained using the PPPM method. Hence, we recommend to give a special attention to this aspect for heterogeneous systems, because it may lead to a qualitative difference in the interfacial structure, which cannot be corrected by any post-simulation routine, and because it may bias comparison of force field predictions with experiments. Currently, many MD softwares provide a long-range implementation of dispersion interactions, which does not involve a large computational cost or can even speed up the calculation \cite{isele2013reconsidering}. Therefore, we suggest here to use an Ewald based method for the long-range dispersion term when dealing with electrolyte solutions at liquid-gas interfaces.

Using such a long-range solver, we reconsidered the impact of the ionic charge rescaling procedure implemented in recent non-polarizable force fields of aqueous electrolytes on surface tension and liquid-vapor interfacial structure. 
With an ECCR force field, we obtained a linear gain of surface tension with respect to ionic concentration, close to the experimental value. We then showed that the charge rescaling factor has a dramatic impact on the local structure in this saline water interfacial system. In particular, we found that a charge rescaling based on short-range arguments ($q = \pm 0.75 \, e$) lead to a positive surface excess while a negative value is expected. We explained that the long-range-motivated charge rescaling factor ($q = \pm 0.85 \, e$) should be preferred for heterogenous systems since this correction apply also to the image charge potential acting on the ions at interfaces with a dielectric contrast.
Finally, we showed that the rescaling of LJ parameters in the ECCR approach, while originally tuned to capture local first neighbor ion-water and ion-ion interactions in bulk, also fairly predicted the formation of an ionic double layer, consistently with polarizable force field results.

We hope that more ECCR models with a charge rescaling factor based on the long-range argument will be established: even if extra work is needed regarding the Van der Waals parameters, significant gains for both homogenous and heterogenous systems can be expected compared to usual non-polarizable force field, and at a computational cost lower than the one of polarizable force fields. 
With that regard, it would be quite interesting to explore in future work how the results obtained here for a NaCl solution would extend to other salts \cite{Zeron2019}, in order to assess further the importance of long-range LJ interactions and the applicability of ECCR models to interfaces.

\section*{Supplemental material}

See the supplemental material for details on: computation of the surface tension of pure water with truncated and untruncated Lennard-Jones interactions; box size convergence; surface tension calculation; surface dividing altitude and surface excess measurement.

\section*{Data Availability Statement}

The data that support the findings of this study are available from the corresponding author upon reasonable request.

\begin{acknowledgments}
LJ acknowledges interesting exchanges with the Madrid group about their model, and fruitful discussions with E. Guillaud and A. Ghoufi. 
This work is supported by the ANR, project ANR-16-CE06-0004-01 NECtAR. LJ is supported by the Institut Universitaire de France. 
This work used the HPC resources from the PSMN mesocenter in Lyon.
\end{acknowledgments}

%



\section*{Supplementary material}

**Supplemental material for "Molecular modeling of aqueous electrolytes at interfaces: effects of long-range dispersion forces and ionic charge rescaling"**

Guillaume Le Breton[1, 2] and Laurent Joly[2, 3, a)]

[1)]*Département de Physique, École Normale Supérieure de Lyon, 46 Allée d'Italie, Lyon Cedex 07, France*

[2)]*Univ Lyon, Univ Claude Bernard Lyon 1, CNRS, Institut Lumière Matière, F-69622, VILLEURBANNE, France*

[3)]*Institut Universitaire de France (IUF)*

---

[a)]Electronic mail: laurent.joly@univ-lyon1.fr

# CONTENTS

## I. SURFACE TENSION OF PURE WATER WITH TRUNCATED AND UNTRUNCATED LENNARD-JONES INTERACTIONS

In this section, we will compute the surface tension of pure water, comparing two methods: one with truncated Lennard-Jones (LJ) interactions complemented with analytical tail corrections, and one with untruncated LJ interactions.

### A. Tail correction for pure water

We considered SPC/E and TIP4P/2005 water, where only oxygen atoms interact through a LJ potential, $V_{\text{LJ}}(r) = 4\varepsilon[(\sigma/r)^{12} - (\sigma/r)^6]$. We could then use an analytical tail correction for the surface tension, proposed initially by Chapela *et al.*[1] and corrected later by Blokhuis *et al.*[2], assuming that the density profile of molecules $\rho(z)$ across the interface can be fitted to a hyperbolic

tangent function:

$$\rho(z) = \frac{1}{2}(\rho_l + \rho_v) - \frac{1}{2}(\rho_l - \rho_v)\tanh\left[\frac{z-z_0}{d}\right], \qquad (1)$$

where $\rho_l$ and $\rho_v$ are the liquid and vapor densities, $z_0$ the interface position and $d$ its thickness. For a LJ cutoff $r_c$, the tail correction can then be written:

$$\gamma_{\text{tail}} = 12\pi\varepsilon\sigma^6(\rho_l - \rho_v)^2 \int_0^1 ds \int_{r_c}^{\infty} dr\, r^{-3}\left(3s^3 - s\right)\coth[sr/d]. \qquad (2)$$

## B. Truncated Lennard-Jones interactions

First, we have computed for the SPC/E and the TIP4P/2005 water models the surface tension of a water film with 1700 water molecules (size $34 \times 34 \times 50\,\text{Å}^3$ in the $x$, $y$ and $z$ directions, with $50\,\text{Å}$ of vacuum above and below the water film) with different cutoffs for the LJ interaction. The systems have been first equilibrated during 300 ps, and the production run lasted for 10 ns. The Coulombic term is treated using a long range Ewald solver in the same way for all the different LJ's cutoff. The water molecule geometry has been made rigid – using the SHAKE algorithm.

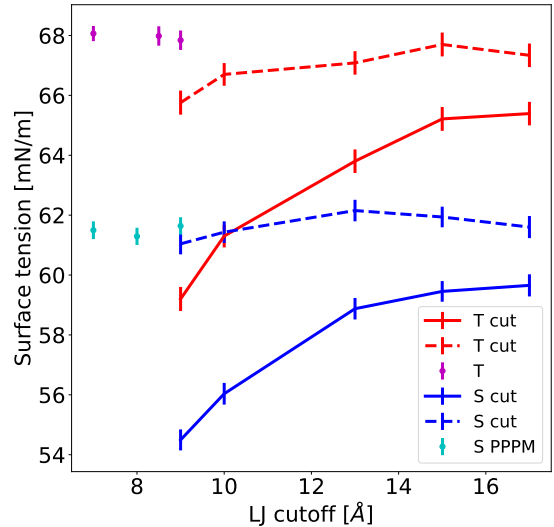

FIG. 1. Comparison between different schemes to compute the Lennard-Jones term: standard cutoff approach with or without analytical tail correction term, and PPPM long-range solver. The experimental surface tension is about 72 mN/m. "T" and "S" stand for TIP4P/2005 and SPC/E water model respectively. The full lines are the raw surface tensions using the cutoff method and the dashed lines include the analytical correction described in the main text. The dots are the raw results using the PPPM method.

3

The results are presented in Fig. 1; the full lines are the averaged raw data and the dashed lines are the tail-corrected values. It is clear that the surface tension is impacted by the used cutoff. The surface tension seems to converge at high cutoff value but the numerical cost is becoming very large. The analytical tail correction provides satisfying results for this system: the corrected surface tension reaches a plateau.

### C. Untruncated Lennard-Jones interactions

We used the particle-particle–particle-mesh (PPPM) method for both the Coulombic and the LJ potential[3,4]. The system used is the same as for the cutoff method. In Fig. 1, one can see that, for both the SPC/E and the TIP4P/2005 water models, using the PPPM method fixes the LJ cutoff dependence. Moreover, the surface tensions obtained are around 2 mN/m larger than the ones obtained using the largest cutoff. This emphasises the fact that the long range part of the LJ potential has an important impact and even a large cutoff misses some relevant interaction for this heterogeneous system. The results obtained between the PPPM method and the cutoff with the tail correction match quantitatively.

## II. BOX SIZE CONVERGENCE

This section presents how the simulation box dimensions have been converged in the in-plane and out-of-plane directions.

### A. In-plane direction

Due to the finite size of the system in the *xy*-dimensions and the periodic boundary conditions, the height fluctuations of the interface can be perturbed for too small in-plane box extension. Indeed, in the reciprocal space, only the large wave-vector are allowed in this case – the most energetic ones. Hence, using periodic boundary conditions in a too small system with a highly constrained interface – large surface tension – can lead to nonphysical behaviour[5]. Therefore, we have tested several box extensions in the *xy*-directions. The simulation time was 5 ns for every systems, and the box length in the *z*-direction was 300 Å. The results are presented in Fig. 2.

For dilute electrolytes (low surface tension), small systems can be used but not for more concentrated ones (high surface tension). A simulation box extension of 34 Å in the *xy*-direction is

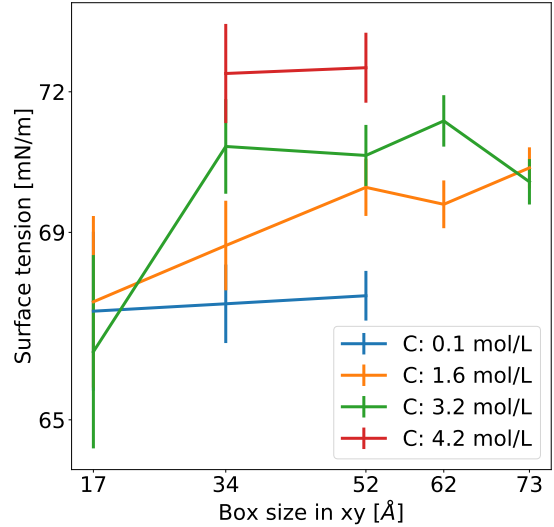

FIG. 2. Box size convergence in the *xy*-directions. Several box extensions along the surface tangential directions were tested for all the studied salt concentrations. The surface tension is used as the observable to converge.

sufficient for all the studied concentrations. This value is close to others found in the literature: 22 Å (Ref. 6) or 40 Å (Ref. 7). Moreover, if the system is too small, the spatial averaging can be less efficient and more time steps are needed. This effect can be clearly seen in Fig. 2 by comparing the 17 Å system and the 34 Å system for the same run duration: the error bar is larger for the smallest system.

### B. Out-of-plane direction

Two issues may also arise along the out-of-plane direction *z*. First, the system we want to describe is made of two "interfaces" areas separated by a "bulk" one. In order to have an idea of the interface extension in the *z*-direction for an aqueous electrolyte, one can use the Debye length.

Hence, to describe a relevant system where the 2 interfaces are separated by a bulk phase, the total extension of the liquid film should be much larger than the Debye length. For the less concentrated systems (0.1 mol/L), the Debye length is around 10 Å, and for the most concentrated one, it drops to less than 1 Å.

In Table I is shown the computed surface tension of a 5 ns production run for several concentra-

5

| film thickness | C: 0.1 | C: 1.6 | C: 3.2 | C: 4.2 |
|---|---|---|---|---|
| 65 Å | 67.9 (0.5) | 69.7 (0.6) | 69.7 (0.7) | 74.5 (0.7) |
| 90 Å | 67.6 (0.2) | 70.0 (0.5) | 70.6 (0.5) | 72.5 (0.6) |

TABLE I. Surface tension [mN.m$^{-1}$] for different salt concentrations $C$ [mol.L$^{-1}$] and liquid film thicknesses [Å]. The the surface tension error is presented between parenthesis next to its average value.

tions and two liquid extensions: approximately 65 Å and 90 Å. No significant difference is found between the two bulk liquid extensions. We have chosen to keep this large 90 Å extension of the liquid phase in order to have more ions at low concentration – 8 Na$^+$ and 8 Cl$^-$ for 3500 water molecule for the lowest concentration in this configuration. This extension is larger than the ones used in some other works[6–8] because of this low ionic concentration.

Second, the system shall not interact with its periodic images along the $z$-axis. Here we used a typical rule of thumb, with a box extension in the $z$ direction that was three times the liquid film thickness. For instance, Ref. 9 has the same water and box simulation $z$-extension.

## III. SURFACE TENSION CALCULATION

This section describes how the surface tension has been obtained and analysed. The surface tension measurement is based on a mechanical description[5].

### A. Mechanical route

The surface tension is calculated using the difference of the normal $p_N(z)$ and tangential $p_T(z)$ components of the local pressure:

$$\gamma = \int_{z_{\min}}^{z_{\max}} [p_N(z) - p_T(z)] \, dz.$$

In the liquid film configuration we considered, the surface tension can be rewritten:

$$\gamma = \frac{L_z}{2} [P_N - P_T] = \frac{L_z}{2} \left[ P_z - \frac{1}{2}(P_x + P_y) \right],$$

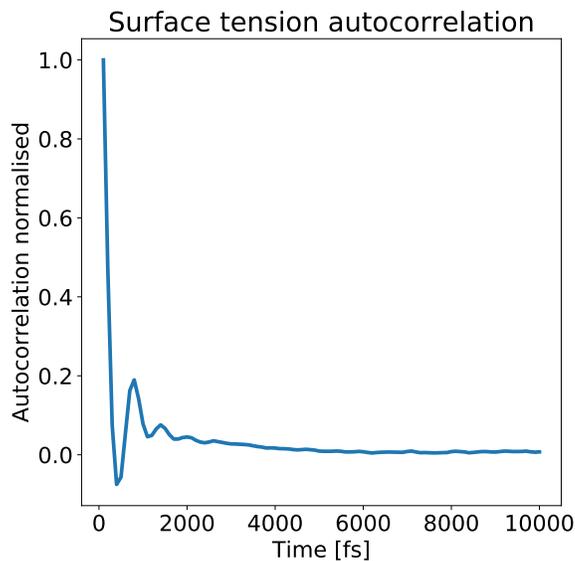

FIG. 3. The surface tension auto-correlation function for TIP4P/2005 pure water system. Each measure is separated by 10 time steps, 20 fs. A sufficient time needed to get independent measurement can be 5000 fs.

where $P_i$ is the averaged pressure over the system at a given time – the $\frac{1}{2}$ is due to the fact that there are 2 interfaces. The pressure is computed using an internal `LAMMPS` calculation, based on a virial definition[10].

## B.  Correlation length and error bars

Simulations have been run for several nanoseconds and the error bars shown have been calculated using a temporal independent set of surface tensions. To do so, we computed the autocorrelation function of the time evolution of the surface tension in a pure TIP4/2005 water system, see Fig. 3. The surface tension was computed every 10 time step – 20 fs. Using this autocorrelation function, we can conclude that surface tension measurements separated by approximately 5 ps are sufficiently independent. Therefore, during this work we have built sets of surface tension obtained every 10 ps from an average over 500 consecutive surface tension values – each of them separated by 10 time steps. Then, the standard deviation is computed assuming this new set is composed of independent measurements. Finally, the error bar amplitude plotted for the surface tension is two times this standard deviation to get a confidence interval of 95% within a gaussian

distribution assumption – confirmed by looking at the surface tension distribution of these kinds of set. We have also tested that these systems were at equilibrium before the production runs, and that 100 ns of production run lead to reproducible results – typical production run time among the literature, for instance similar to Ref. 7.

## IV. SURFACE DIVIDING ALTITUDE AND SURFACE EXCESS MEASUREMENT

In this section, we will describe how the surface excess (SE) is measured using water and ionic density profiles.

The SE is a way to quantify how the solute density behaves at the interface edge. This quantity, given in particle number per nm$^2$, is defined by:

$$\Gamma = \int_{-\infty}^{z_G} \left[ \rho^{\text{ion}}(z) - \rho_b^{\text{ion}} \right] dz + \int_{z_G}^{+\infty} \rho^{\text{ion}}(z) dz$$

Where $\rho^{\text{ion}}(z)$ is the ionic density along the out-of-plane direction, $\rho_b^{\text{ion}}$ the ionic "bulk" density and $z_G$ the Gibbs dividing altitude. To determine the Gibbs dividing altitude, the water distribution is used and two different definitions have been tested. The first is provided by the hyperbolic tangent fit of the water density profile across the interface, see Eq. 1. The second is obtained through:

$$\int_{-\infty}^{+\infty} \rho^{\text{wat}}(z) dz = \int_{-z_G}^{+z_G} \rho_b^{\text{wat}} dz,$$

where $\rho^{\text{wat}}(z)$ and $\rho_b^{\text{wat}}$ the water density along the out-of-plane axis and in the bulk phase respectively. As pointed by Ref. 8, we also found quantitative agreement between these two definitions. In order to quantify the uncertainty for the SE, we have chosen to major and minor the calculated bulk ionic concentration $\rho_b^{\text{ion}}$. This interval is then used to get the error bar of the SE plotted in the Fig. 3 of the main text.



\end{document}